\documentclass[review]{elsarticle}

\usepackage{lineno,hyperref,amsmath,xcolor}

\journal{Journal of \LaTeX\ Templates}

\begin{document}

\begin{frontmatter}

\title{\textcolor{black}{Compressed ghost edge imaging}}

\author[mymainaddress,address]{Hui Guo}
\author[mymainaddress]{Le Wang}
\author[mymainaddress]{Shengmei Zhao \corref{mycorrespondingauthor}}
\cortext[mycorrespondingauthor]{Corresponding author}
\ead{zhaosm@njupt.edu.cn}
\address[mymainaddress]{Institute of Signal Processing and Transmission, Nanjing University of Posts and Telecommunications, Nanjing, Jiangsu 210003, China}
\address[address]{College of Information Engineering, Fuyang Normal University, Fuyang, Anhui 236037, China}

\begin{abstract}
In this paper, we propose an advanced framework of ghost edge imaging, named compressed ghost edge imaging (CGEI). In the scheme, a set of structured speckle patterns with pixel shifting are illuminated on an unknown object, and the output is collected by a bucket detector without any spatial resolution. By using compressed sensing algorithm, we obtain the horizontal and vertical edge information of the unknown object with the bucket detector detection results and the known structured speckle patterns. The edge is finally constructed by the two-dimentional edge information. The experimental and numerical simulations results show that the proposed scheme has a higher quality and reduces the number of measurements, in comparison with the existed edge detection schemes based on ghost imaging.
\end{abstract}


\end{frontmatter}

Ghost imaging (GI), also termed single-pixel imaging, is a novel optical imaging technique which has received great attention recently [1-6]. There are two spatially correlated optical beams in a GI system. One beam, called the object beam, illuminates an unknown object and is then collected by a bucket detector without any spatial resolution. The other beam, named the reference beam, never interacts with the object and is detected by a spatially resolving detector. A ghost imaging is retrieved by correlating the bucket signal and the reference sigal, but not either one alone. Compared with the traditional imaging methods, GI can be used to retrieve the image of the object in various optical harsh or noisy
environments[7-11]. To improve the imaging quality of GI, various GI methods were proposed, such as iterative GI[12], differential GI[13], compressive GI[14], normalized GI[15], sinusoidal GI[16,17] and so on[18-20].\\
\indent Edge detection tests the object edge consisting of a dramatic change in image processing. It has extensive usage in image segmentation, target recognition, and computer vision[21,22]. In traditional edge detection methods, the object needs to be imaged at first, and the edge information can be obtained by the corresponding edge operator. However, in harsh or noisy environments, the imaging of the object is difficult to achieve, so the edge detection algorithm cannot be implemented. Owing to the special properties of GI mentioned above, edge detection based on GI can solve the problem of disturbance in the imaging optical path and has an advantage in edge detection. In recent years, GI based edge detection has achieved some results [23-27]. In [23], a gradient GI (GGI) was proposed to achieve the edges of an unknown object directly. However, it is a problem to choose a proper gradient angle based on the prior knowledge of the object in this method. Subsequently, the speckle-shifting ghost imaging (SSGI) was introduced to achieve the edges of an unknown object without any other prior knowledge of the object[24]. Meanwhile, a subpixel-speckle-shifting ghost imaging was proposed which can enhancing the resolutions of the edge detection with low resolution speckle patterns[25]. In [26], authors presented structured intensity patterns to get the edge of an object directly by the data detected in CGI. In [27], special sinusoidal patterns were designed to get the edge of the unknown object with an improvement in signal-to-noise ratio (SNR) in the frequency domain. However, the number of measurements of these schemes is still large and the quality of edge detection results still need to be improved. \\
\indent On the other hand, compressed sensing (CS) method was introduced into GI to obtain a higher resolution image of the object by exploiting the redundancy in the structure of the images to reduce the number of measurements required for exact reconstruction[28-31]. Therefore, The GI based on compressed sensing that enables the reconstruction of an N-pixel image from much less than N measurements, overcomes the limitation of Nyquist sampling theorem, and greatly reduces the acquisition time and requisite samples[32-34].\\
\indent In the paper, we propose an edge detection scheme with CS technique, named compressed ghost edge imaging (CGEI). In the scheme, the special random patterns with the characteristic of different speckle-shifting are first designed. With CS technique, the horizontal and vertical edge information with high quality could be obtained directly with the bucket detector detection results  and the structured illuminations. Lastly, the global edge of the unknown object is constructed with the two-dimensional edge information.\\
\begin{center}
\includegraphics[scale=0.45]{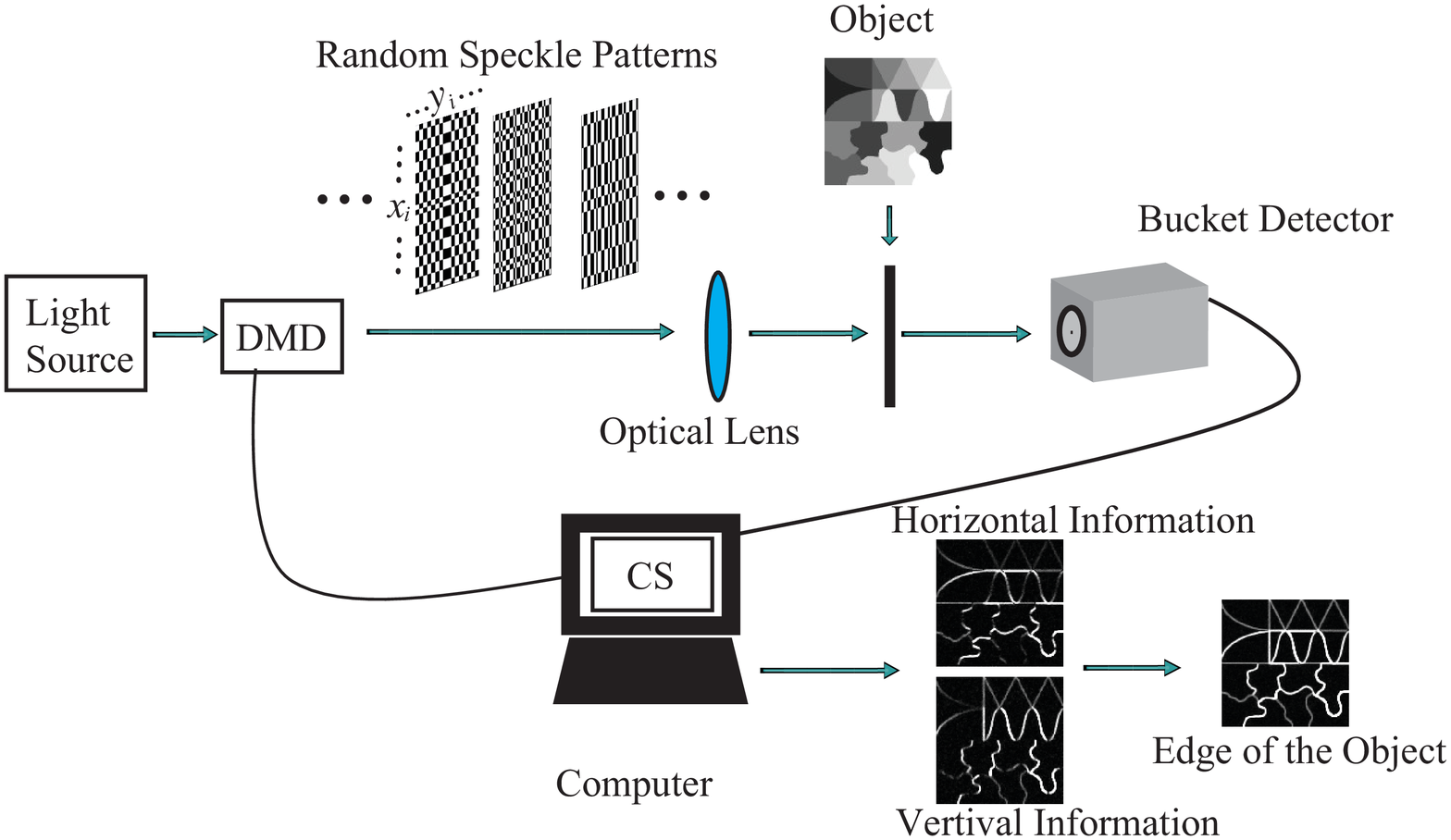}
\end{center}
\vskip 0.5cm \small \rm Fig.\hspace{0.1cm}1. A schematic diagram of the CGEI scheme
\vskip 0.8cm \noindent \normalsize
\indent Fig.\hspace{0.1cm}1 shows the schematic diagram of the CGEI scheme. The light is modulated by a digital micro-mirror device (DMD) which is controlled by a computer to produce the speckle patterns $S_k(x_i,y_j)$, $k=1,2, \dotsm, M$, $M$ is the number of measurements, $x_i,y_j$ are the spatial coordinates. The bucket detector measures the total light transmitted the object $T(x_i,y_j)$, and the output signal is detected as,
\begin{align}
y_k=\sum_{x_i}\sum_{y_j}S_k(x_i,y_j)T(x_i,y_j) \label{con:inventoryflow1}.
\end{align}
Here, $S_k(x_i,y_j)$ can be pre-designed. The image of the object can be obtained by the second-order correlation,
\begin{align}
T(x_i,y_j)=\big<S_k(x_i,y_j)\cdot y_k\big>-\big<S_k(x_i,y_j)\big>\big<y_k\big> \label{con:inventoryflow2}.
\end{align}
Here $\big<\cdot\big>$ denotes ensemble average. \\

\indent Speckle-shifting, caused by edge operator (such as Gradient vector and Sobel operator etc.),  makes GI achieve the edge of an unknown object directly [23,24]. Here, we take the Sobel operator as an example to introduce the principle of CGEI. To realize edge detection, several speckle groups are divided and they are related as follows
\begin{align} \begin{split}
S_k(x_i,y_j)& =S_k^1(x_{i-1},y_{j-1})=S_k^2(x_{i-1},y_{j})\\
            & =S_k^3(x_{i-1},y_{j+1})=S_k^4(x_{i},y_{j-1})\\
            & =S_k^5(x_{i},y_{j+1})=S_k^6(x_{i+1},y_{j-1})\\
            & =S_k^7(x_{i+1},y_{j})=S_k^8(x_{i+1},y_{j+1}),\\ \label{con:inventoryflow3}
\end{split} \end{align}
where $S_k^l, l=1, 2, \dotsm, 8$, represents the $l$th group of shifted speckle patterns. By using the Sobel operator property and Eq. (\ref{con:inventoryflow1}), one obtains the horizontal measurements as follows:
\begin{align} \begin{split}
\nabla y_k^h &=\sum_{x_i}\sum_{y_j}S_k^1(x_{i},y_{j})T(x_i,y_j)\\
           &\quad +2\times\sum_{x_i}\sum_{y_j}S_k^2(x_{i},y_{j})T(x_i,y_j)\\
           &\quad +\sum_{x_i}\sum_{y_j}S_k^3(x_{i},y_{j})T(x_i,y_j)\\
           &\quad -\sum_{x_i}\sum_{y_j}S_k^6(x_{i},y_{j})T(x_i,y_j)\\
           &\quad -2\times\sum_{x_i}\sum_{y_j}S_k^7(x_{i},y_{j})T(x_i,y_j)\\
           &\quad -\sum_{x_i}\sum_{y_j}S_k^8(x_{i},y_{j})T(x_i,y_j)\\
           &= \sum_{x_i}\sum_{y_j}S_k(x_i,y_j)(T(x_{i-1},y_{j-1})+2\times T(x_{i-1},y_{j})\\
           &\quad +T(x_{i-1},y_{j+1})-T(x_{i+1},y_{j-1})\\
           &\quad -2\times T(x_{i+1},y_{j})-T(x_{i+1},y_{j+1}))\\
           &=\sum_{x_i}\sum_{y_j}S_k(x_i,y_j)\nabla_h^{S}T(x_i,y_j),\\
\end{split} \end{align}
here, $\nabla_h^{S}T(x_i,y_j)$  represent the horizontal edges of the unknown object using Sobel operator. Using Eq. (\ref{con:inventoryflow2}), the horizontal edges of the object as
\begin{align}\begin{split}
\nabla_h^{S}T(x_i,y_j)
&=\big<S_k(x_i,y_j)\cdot \nabla y_k^h\big>-  \\
&\quad \big<S_k(x_i,y_j)\big>\big<\nabla y_k^h\big>.
\end{split}\end{align}
Similarly, the vertical edge of the object is obtained
\begin{align}\begin{split}
\nabla_v^{S}T(x_i,y_j)
&=\big<S_k(x_i,y_j)\cdot \nabla y_k^v\big>- \\
&\quad \big<S_k(x_i,y_j)\big>\big<\nabla y_k^v\big>,
\end{split}\end{align}
where, $\nabla y_k^v$ is the vertical measurements and $\nabla_v^{S}T(x_i,y_j)$ represents the vertical edges of the unknown object using Sobel operator. Finally, the edge of the object will be obtained by
\begin{align}\begin{split}
T_{edge}=\sqrt {(\nabla_h^{S}T(x_i,y_j))^2 +(\nabla_v^{S}T(x_i,y_j))^2}.
\end{split}\end{align}\\
\indent However, the edges reconstruction of the unknown object requires too many measurements from above method. To solve the problem, we use compressed sensing technique. Here, the total variation minimization by augmented Lagrangian and alternating direction algorithms (TVAL3) [33,34] is adopted. The speckle field of the $k$th sample is recorded as $S_k(x_i,y_j)$. The indice $x_i$ represents the horizontal pixel coordinates and $i=1, 2, \dotsm, m$, the indice $y_j$ represents the vertical pixel coordinates and $j=1, 2, \dotsm, n$. The indices $k$ is the sampling frame index and $M$ is the total speckle patterns number. Then, each of the speckle intensity $S_k(x_i,y_j)$ is reshaped as a row vector $\phi_k$ of size $1\times N$, where $N=m\times n$.
\begin{align}\begin{split}
\phi_k & =\big [S_k(x_1,y_1)\quad \dotsm \quad S_k(x_1,y_n) \quad S_k(x_2,y_1) \quad \dotsm\\
       &\qquad S_k(x_2,y_n) \quad \dotsm \quad S_k(x_m,y_1) \quad \dotsm \quad S_k(x_m,y_n) \big ],
\end{split}
\end{align}
after $M$ samples, we can get a $M\times N$  samples array recorded as $A$, and it can be written as the following matrix
\begin{align}
\begin{split}
A& =\left[
\begin{matrix}
\phi_1\\
\phi_2\\
\vdots\\
\phi_M\\
\end{matrix}
\right]\\
& =\left[
\begin{matrix}
S_1(x_1,y_1)&S_1(x_1,y_2)&\dotsm &S_1(x_m,y_n)\\
S_2(x_1,y_1)&S_2(x_1,y_2)&\dotsm &S_2(x_m,y_n)\\
\vdots &\vdots &\ddots &\vdots\\
S_M(x_1,y_1)&S_M(x_1,y_2)&\dotsm&S_M(x_m,y_n)\\
\end{matrix}
\right],
\end{split}
\end{align}\\
meanwhile, $M$ speckle patterns with $l$th group shifting divide the results from the bucket detector into two groups. The value of bucket detector with horizontal information of the unknown object can be arranged as a $M\times 1$ column vector $Y^h$ i.e., $Y^h=\big [\nabla y^h_1\quad\nabla y^h_2\quad\dotsm \quad\nabla y^h_k\quad\dotsm\quad\nabla y^h_M\big ]^T$. Take $\nabla y^h_k$ for example, it can be obtained by
\begin{align}\begin{split}
\nabla y^h_k=y^1_k+ 2\times y^2_k+y^3_k-y^6_k-2\times y^7_k-y^8_k,
\end{split}
\end{align}\\
where $y^l_k, l=1,2,3,6,7,8$ are the bucket values corresponding $S^l_k, l=1,2,3,6,7,8$ passing through the unknown object. Then, if we denote the edge information of the unknown object in horizontal direction as an $N$ dimensional column vector $X^h(N\times 1)$. Therefore, $X^h(x_i,y_j)$ could be reconstructed as a solution
\begin{align}\begin{split}
min \Vert DX^h \Vert_1+ \frac\mu2 \Vert Y^h-AX^h \Vert_2^2,\label{con:inventoryflow10}
\end{split}
\end{align}\\
and we can get vertical information $X^v(x_i,y_j)$ in the same way. Here, $\mu$ is a nonnegative parameter, the sparse transform $D$ is usually exploited a set of fixed bases, such as discrete cosine transform and wavelet, $ \Vert \cdot \Vert_1$ and $\Vert \cdot \Vert_2$ stand for the $l_1$ norm and $l_2$ norm, respectively.\\
\indent To compare the quality of the edge detection quantitatively, the signal-to-noise ratio (SNR) is used as an objective evaluation, which is determined from the following definition[23-27].
\begin{align}\begin{split}
SNR=\frac{mean(T_{edge})-mean(T_{back})}{{(var(T_{back}))}^{0.5}},\label{con:inventoryflow12}
\end{split}
\end{align}\\
where, $T_{edge}$ and $T_{back}$ are the intensity values of the edge detection results and background region, respectively, $mean(\cdot )$ represents the average value, and $var(\cdot )$ denotes the variance value. At the same time, we introduce the definition of compression ratio
\begin{align}\begin{split}
Compression \ ratio=\frac{M}{m\times n}=\frac{M^{'}}{m\times n\times l},
\end{split}
\end{align}\\
where, $M$ is the number of the speckle patterns, $M^{'}$ is the number of the measurement of the bucket detector, $m$ and $n$ represent the horizontal and vertical dimensions of the object, respectively, and $l$ is the number of the speckle patterns one-pixel shifting. The $l$ for these algorithms using Gradient vector and Sobel operator are 2 and 8, respectively.\\
\indent For comparison, numerical simulations and experiments are taken for the gray-scale object. Because of the randomness of speckle patterns intensity distribution, we present the reconstructed image over 10 times in all of the following simulations and experiments, respectively, and the speckle patterns used in different methods of the same object is completely consistent. The simulation and experimental results (SNR) selected in the following paragraphs is the 5th of 10 measurements of SNRs arranged in small to large order, and we set $\mu$ equal $2^{12}$ in Eq.(\ref{con:inventoryflow10}) as a coefficient to balance the regularization and the data fidelity.\\
\indent To evaluate the effectiveness of CGEI, we start with numerical simulations. The edge detection results of GGI with a Gradient vector of $\varphi = 45^\circ$ (GGI-$45^\circ$), SSGI using Sobel operator (SSGI-So), CEGI with a gradient of $\varphi = 45^\circ$ (CEGI-$45^\circ$), and CEGI using Sobel operator (CGEI-So) are summarized in Fig.\hspace{0.1cm}2. There are 6554 speckle patterns (Compression ratio: 0.4) modulated by DMD to illuminate onto the gray-scale objects ($128\times128$ pixels). In the GGI scheme, GGI-$45^\circ$ are implemented with random patterns with two one-pixel shifting, so the measurement of the bucket detector is twice the number of the random patterns (total 13108 measurements). Similarly, the bucket detector measurements of SSGI-So with eight one-pixel shifting eight times the number of random patterns (total 52432 measurements) respectively. Meanwhile, the number of bucket measurements of CGEI-$45^\circ$, and CGEI-So are 13108, and 52432, respectively. The results in Fig.\hspace{0.1cm}2 show that CGEI enhances of the ability to recognize the targets than GGI-$45^\circ$ and SSGI-So. For ``Picture1'', the SNR of edge detection using CGEI-$45^\circ$, CGEI-So, GGI-$45^\circ$ and SSGI-So are 5.2383, 10.0509, 0.6544, 1.1592, respectively, and the use of CGEI has enhanced SNR more than eight times. In addition, we also use ``Picture2'' as the target and the simulation results are summarized in Fig.\hspace{0.1cm}2. Compared with the results of different schemes, we can see our scheme has a better SNR
performance.\\
\begin{center}
\includegraphics[scale=0.4]{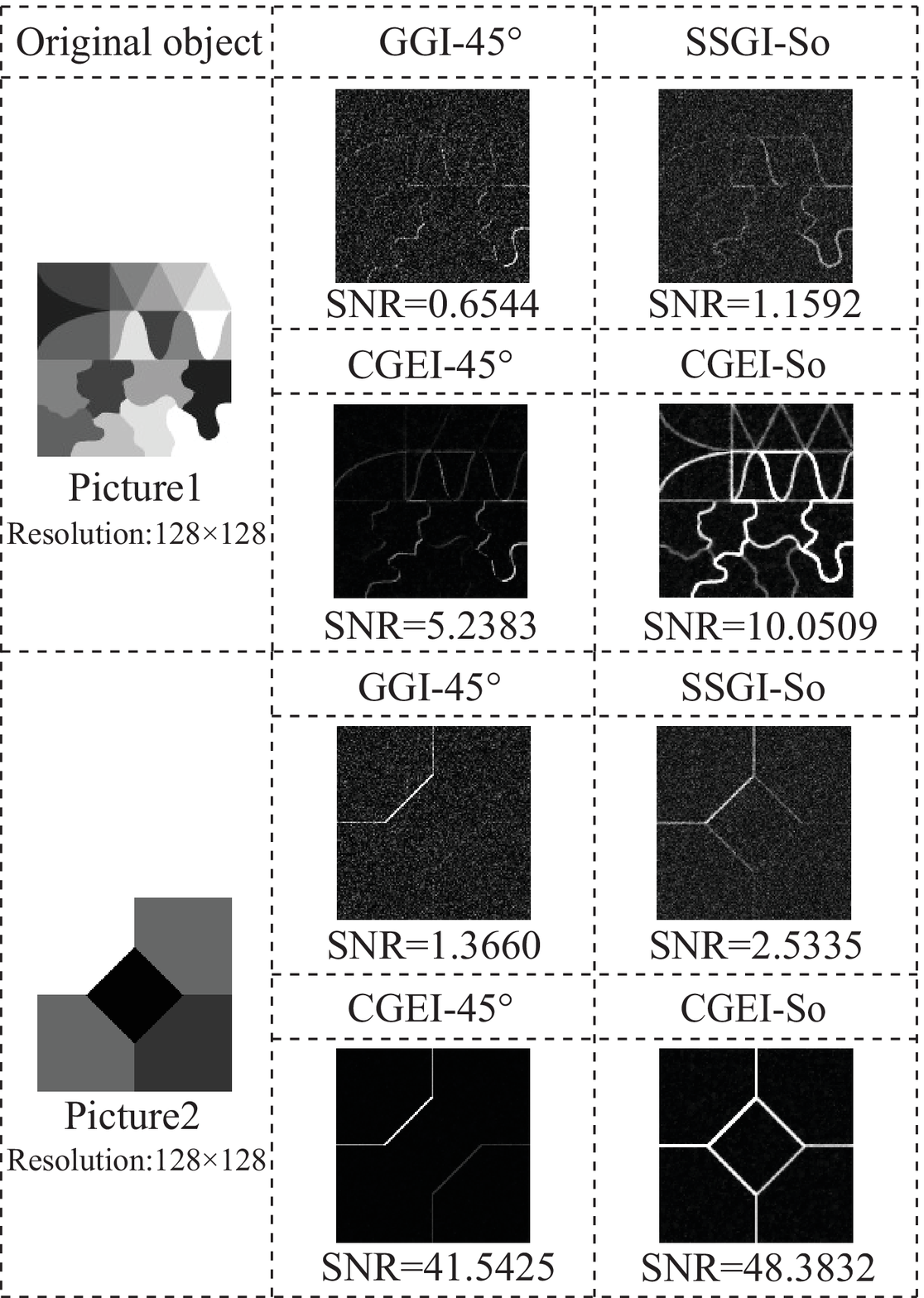}
\end{center}
\vskip 0.5cm \small \rm Fig.\hspace{0.1cm}2. The numerical simulation results of the unknown objects using different schemes, where SNRs are presented together.
\vskip 0.8cm \noindent \normalsize
\begin{center}
\includegraphics[scale=0.3]{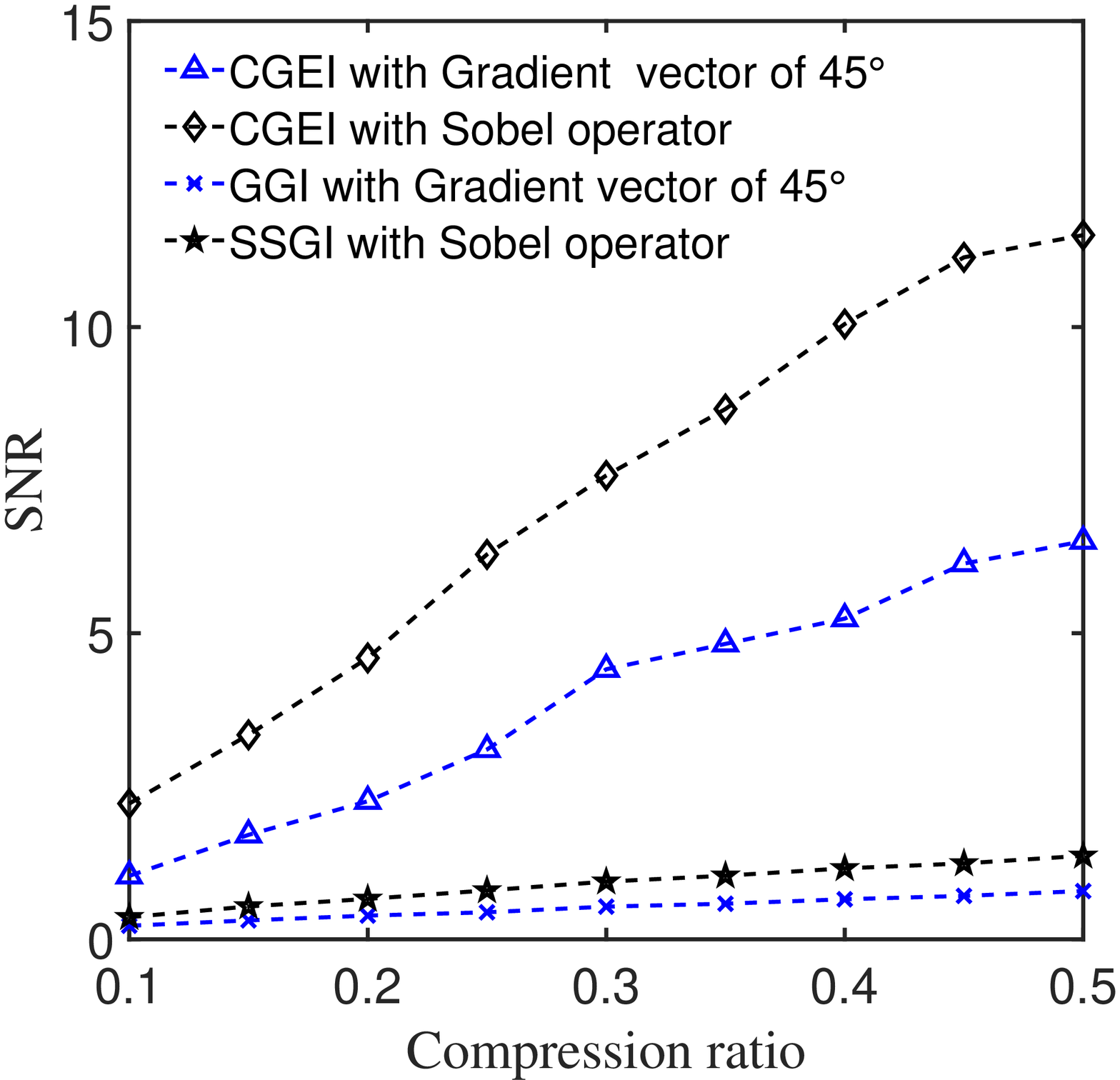}
\end{center}
\vskip 0.5cm \small \rm Fig.\hspace{0.1cm}3. The numerical simulation results (SNR) of different schemes with different compression ratio.
\vskip 0.8cm \noindent \normalsize
\indent Next, we simulate GGI-$45^\circ$, SSGI-So, CGEI-$45^\circ$, and CGEI-So with different compression ratio. The results (SNR) of different algorithms, as a function of measurements, where ``Picture1'' is the target, are expressed in Fig.\hspace{0.1cm}3. It is intuitively seen that using CGEI perform much better than GGI-$45^\circ$ and SSGI-So, regardless of the compression ratio.\\
\indent There are a lot of various noise in real applications. According to Eq. (\ref{con:inventoryflow1}), Added White Gaussian Noise (AWGN) is considered to simulate in the bucket detection. The SNR of the bucket detector $SNR_{BD}$ is defined as
\begin{align}\begin{split}
SNR_{BD}=10log_{10}\frac{Power_S}{Power_N},\label{con:inventoryflow13}
\end{split}
\end{align}
 where $Power_S$ is the signal power collected by the bucket detector and $Power_N$ is the power of AWGN with zero means imposed on the bucket detector. The results (SNRs) of different algorithms to ``Picture1'', where the number of speckle patterns is set to 4915 (Compression ratio: 0.3), are shown in Fig.\hspace{0.1cm}4. From Fig.\hspace{0.1cm}4, we can see that with the increase of the power of the AWGN, the SNRs of all the edge detection algorithms decrease. The SNR of CGEI outperforms that of GGI-$45^\circ$ and SSGI-So when $SNR_{BD}$ is higher than 9dB.\\
\begin{center}
\includegraphics[scale=0.3]{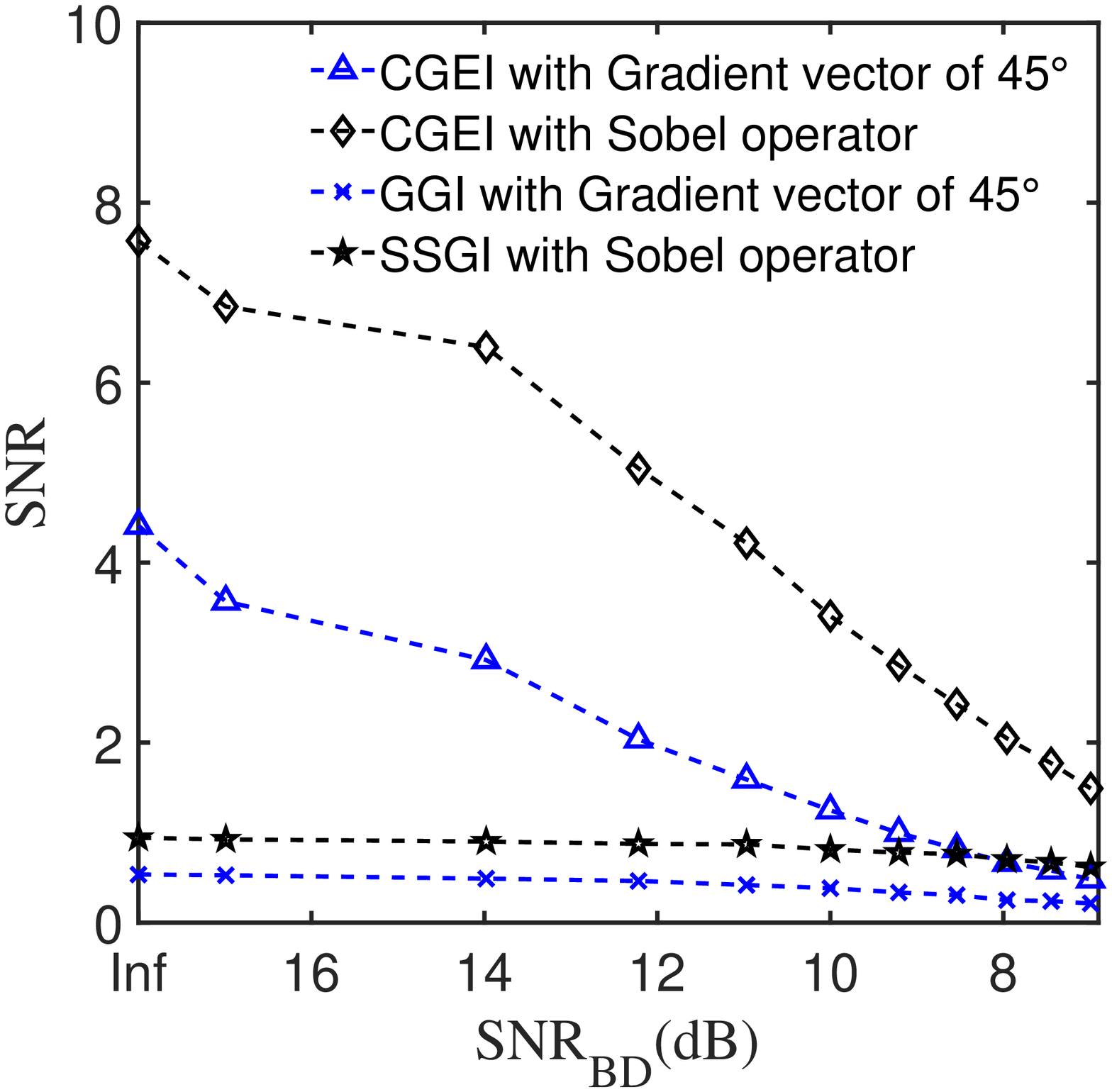}
\end{center}
\vskip 0.5cm \small \rm Fig.\hspace{0.1cm}4. The $\rm SNR$ of edge detection against $\rm SNR_{BD}$ with different schemes.
\vskip 0.8cm \noindent \normalsize\\
\indent Moreover, we perform our scheme experimentally. The experimental system of CGEI is shown in Fig.\hspace{0.1cm}5. One red led is used as the source and the light collimated by lens L1 to focuse on a Digital micromirror device (DMD TIDLPC350). Then the DMD controlled by a computer to modulate the light to generate the random speckle patterns, $S_k^l(x_i,y_j), k=1, 2, \dotsm, M$. Later, the beams with the random speckle patterns are projected onto an object by lens L2. The model of a rabbit (as the gray-scale object), is used as the unknown object in Fig.\hspace{0.1cm}5. The transmitted light carrying the object information is collected by an imaging lens L3 onto a bucket detector (Thorlabs PMM02) to complete the measurement. A pair of detection results $\nabla y_k^h$ and $\nabla y_k^v$ are recorded by the computer. Then, the edges of the object in horizontal and vertical direction are extracted by using the TVAL3 algorithm. Finally, we get the edge of the unknown object. \\
\indent The experimental results were summarized in Fig.\hspace{0.1cm}6. The number of the speckle patterns are set to 2000 (Compression ratio: 0.49). Similar to our theoretical analysis and simulations, it is clear that the results with CGEI are better than GGI-$45^\circ$ and SSGI-So in real applications.\\
\begin{center}
\includegraphics[scale=0.35]{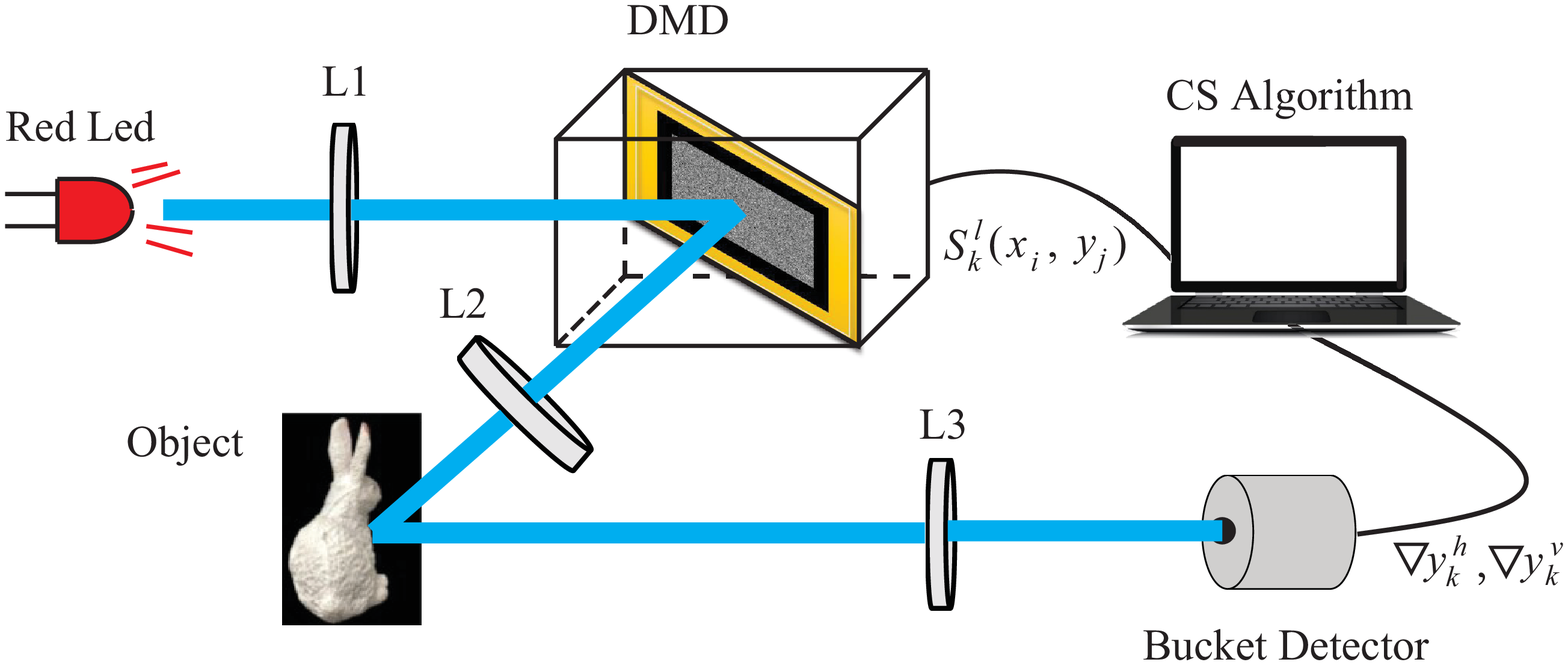}
\end{center}
\vskip 0.5cm \small \rm Fig.\hspace{0.1cm}5. A schematic diagram of the CGEI scheme experimental system.
\vskip 0.8cm \noindent \normalsize\\
\begin{center}
\includegraphics[scale=0.5]{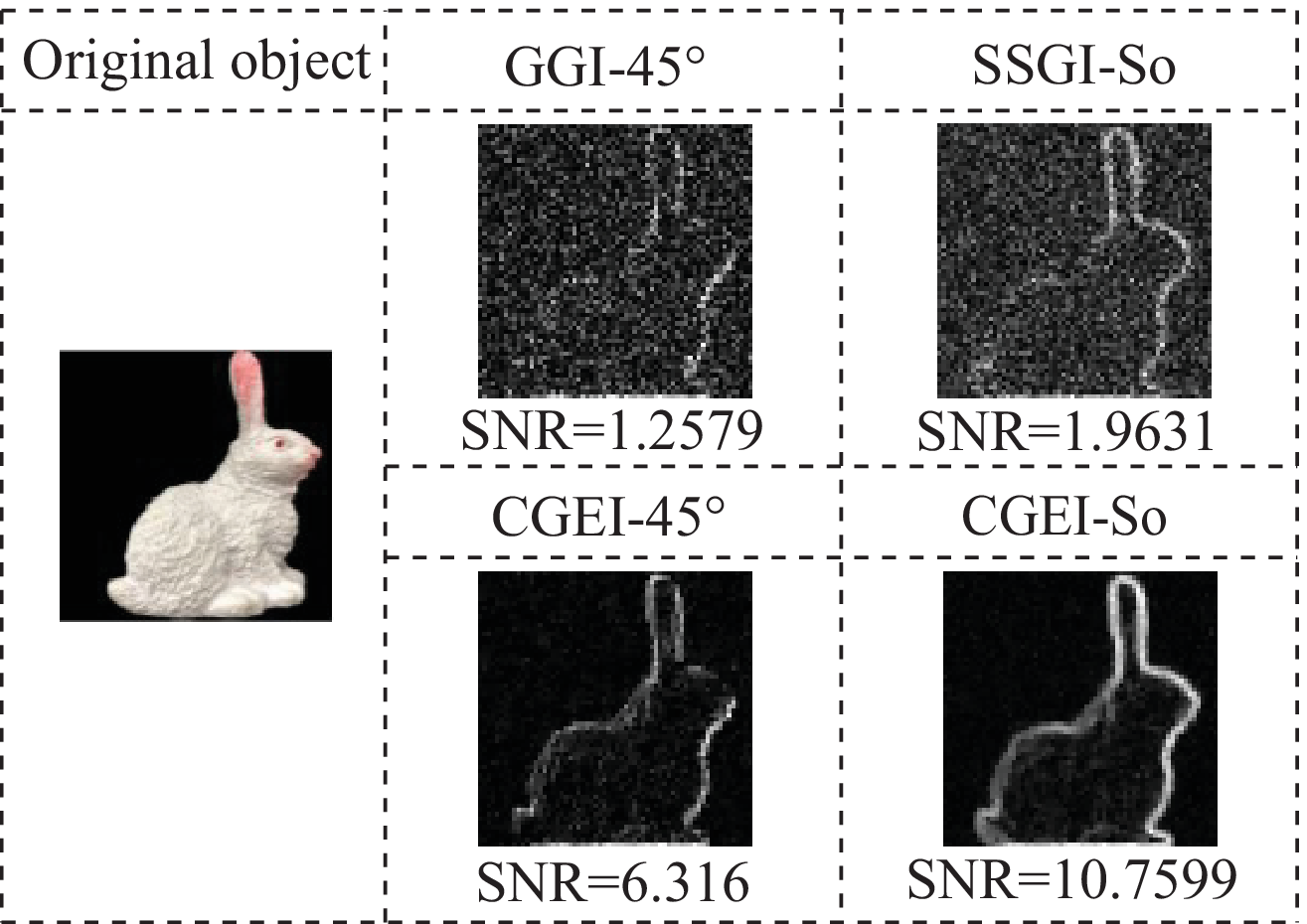}
\end{center}
\vskip 0.5cm \small \rm Fig.\hspace{0.1cm}6. The experimental results of the unknown object using different schemes, where SNRs are presented together and the resolution of results is $64\times64$.
\vskip 0.8cm \noindent \normalsize\\
\begin{center}
\includegraphics[scale=0.3]{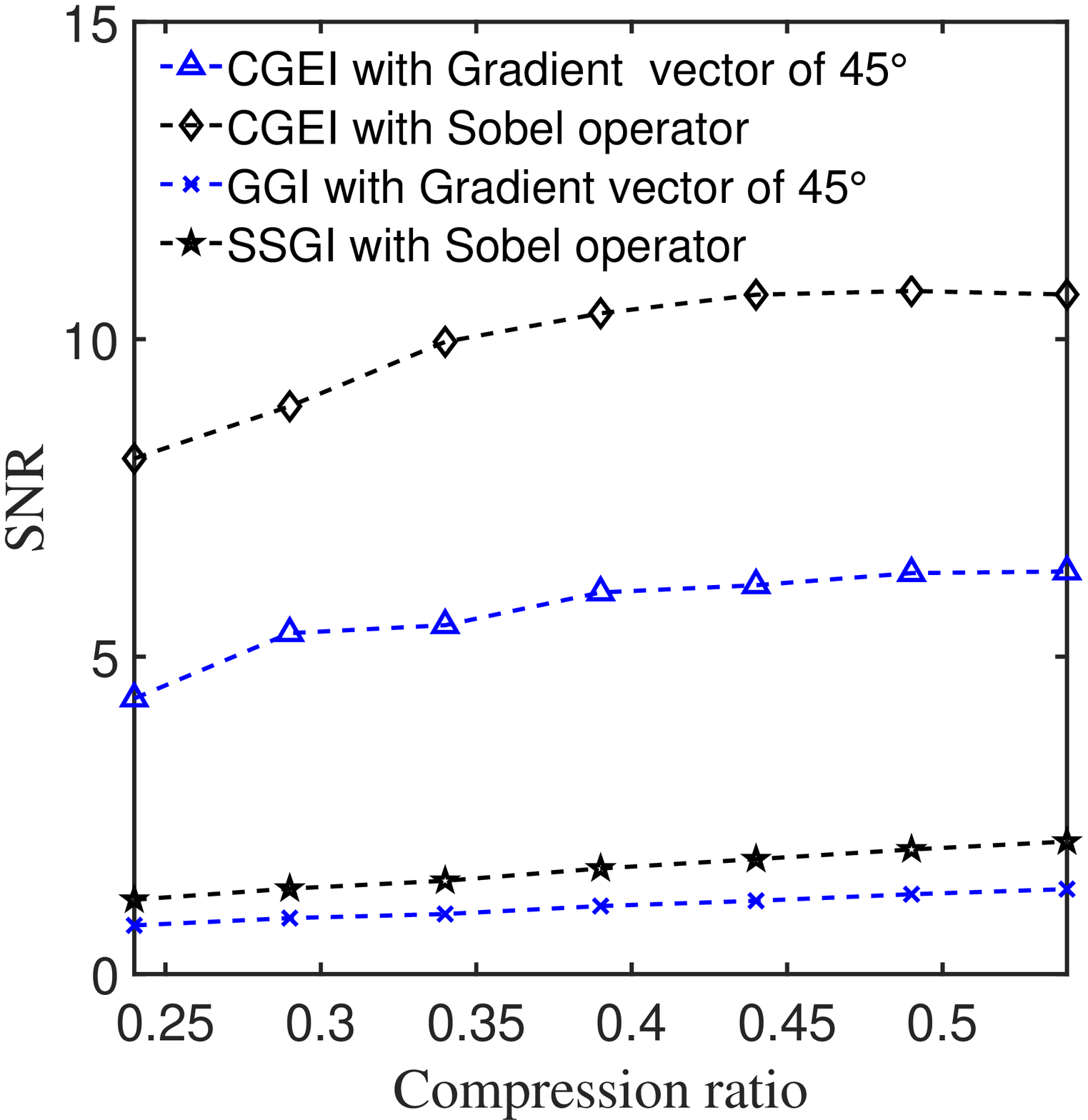}
\end{center}
\vskip 0.5cm \small \rm Fig.\hspace{0.1cm}7. The experimental results of the unknown object using different compression ratio, where SNRs are presented together.
\vskip 0.8cm \noindent \normalsize
\indent In addition, we carry out the experimental results using different schemes with different compression ratio and the results are shown in Fig.\hspace{0.1cm}7, where the model of the rabbit is still taken as the target. It can be seen from Fig.\hspace{0.1cm}7, the experimental results show that CGEI outperforms other algorithms.\\
\indent To further illustrate the effectiveness of CGEI, we do the following simulation comparison. In most traditional edge detection approaches the original target is imaged firstly, and then the edge is extracted with certain algorithms such as Gradient vector and Sobel operator. In the simulation we obtain the object image by using the compressed ghost imaging[14], and then use the Gradient vector (CSGI-$45^\circ$) and Sobel operator (CSGI-So) to extract the edge of the object image. The number of the speckle patterns is set to 4915 (Compression ratio: 0.3). It is seen that the SNRs of edge detection of the reconstructed images with CSGI-$45^\circ$ and CSGI-So are lower than that of CGEI-$45^\circ$ and CGEI-So respectively in Fig.\hspace{0.1cm}8, which further illustrate the effectiveness of CGEI. Hence, CGEI can extract the edge information of an unknown object without needing the original image, and dramatically improve the performance of the edge detection.\\
\begin{center}
\includegraphics[scale=0.45]{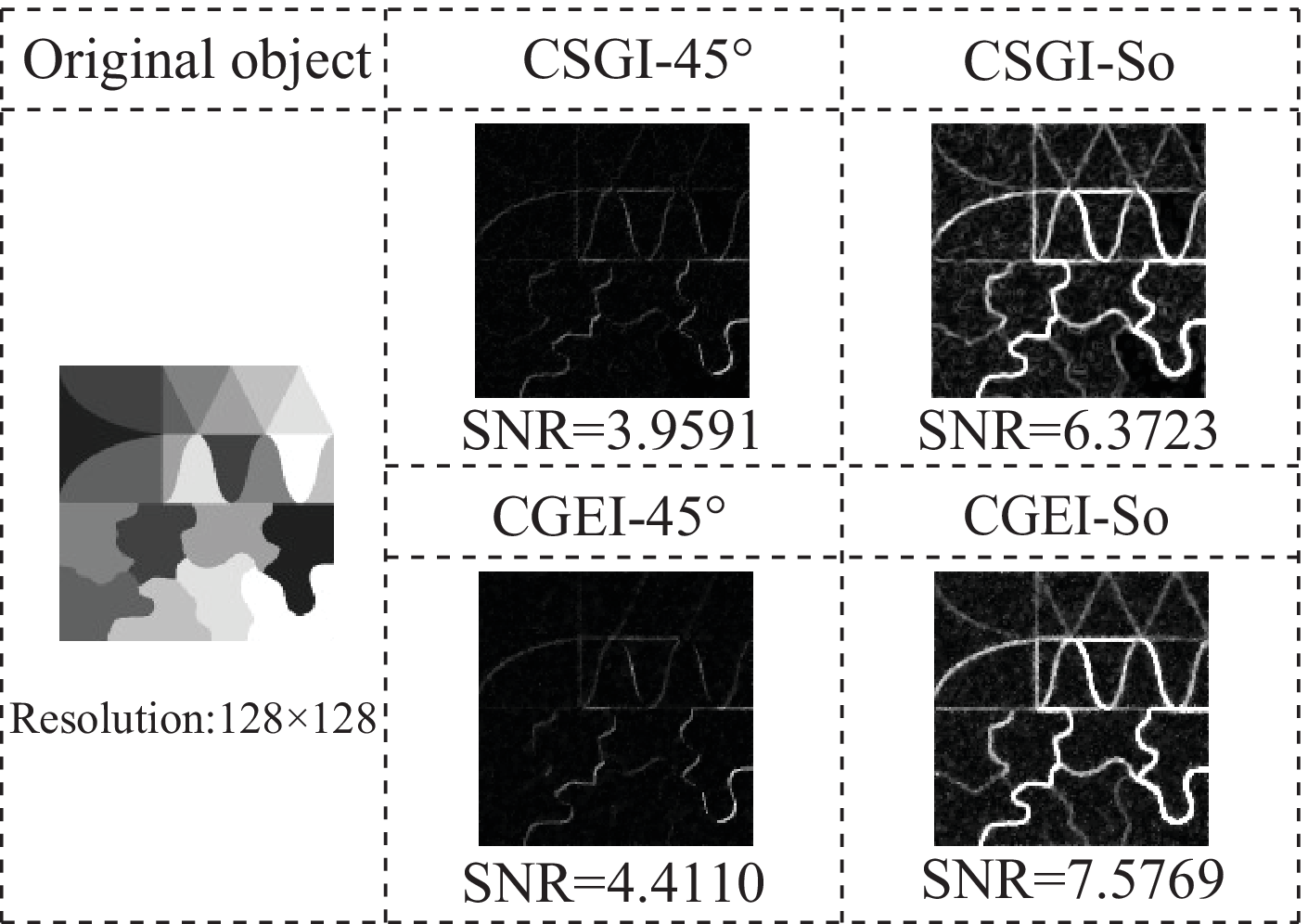}
\end{center}
\vskip 0.5cm \small \rm Fig.\hspace{0.1cm}8. The simulation results (SNRs) of  CSGI-$45^\circ$, CSGI-So, CGEI-$45^\circ$ and CGEI-So.
\vskip 0.8cm \noindent \normalsize
\indent In conclusion, we have proposed an edge detection scheme by using compressed ghost imaging (CGEI) in the paper. We have compared the performance of GGI-$45^\circ$, SSGI-So, CGEI-$45^\circ$ and CGEI-So by numerical simulations and experiments. The results have shown that the imaging quality could be greatly improved by using CGEI comparing with GGI-$45^\circ$, SSGI-So. Moreover, we make a simulation comparison to prove that CGEI-$45^\circ$ and CGEI-So have good performance than CSGI-$45^\circ$ and CSGI-So respectively. Simulation and experiment show that CGEI has a good edge detection effect.\\

\section*{Acknowledgment}
This work was supported by the National Natural Science Foundation of China (Nos. 61871234 and 11847062), the Postgraduate Research \& Practice Innovation Program of Jiangsu Province (No. KYCX18\_0900) and the Natural Science Research Project of Fuyang Normal University (No. 2015FXXZK02)

\end{document}